\documentstyle[twocolumn,aps,epsf,floats]{revtex}
\begin{document}

\draft
\title{Highly anisotropic commensurability oscillations in two-dimensional holes at the GaAs/AlGaAs (311)A interface}

\author{J. B. Yau, J. P. Lu, H. C. Manoharan, and M. Shayegan}
\address{Department of Electrical Engineering, Princeton University, Princeton, New Jersey 08544}
\date{\today}
\maketitle
\begin{abstract}
Measurements of commensurability oscillations in GaAs/AlGaAs two-dimensional (2D) hole systems grown on GaAs (311)A substrates reveal a remarkable anisotropy: the amplitude of the measured commensurability oscillations along the [$\bar{2}$33] direction is about 100 times larger than along [01$\bar{1}$]. For 2D electron systems at similar interfaces, however, we observe nearly isotropic oscillations, suggesting that the anomalous anisotropy is intrinsic to GaAs 2D holes at the (311)A interface.
\end{abstract}

\vspace{0.4cm}
GaAs/AlGaAs heterojunctions and quantum wells, grown on GaAs (311)A substrates and modulation-doped with Si, are of considerable current interest. These structures can contain very high mobility two-dimensional {\it hole\/} systems (2DHSs), and have already shown a number of novel phenomena including signatures of magnetic-field-induced Wigner crystallization\cite{Santos92} and metal-insulator transition at zero magnetic field\cite{Hanein98,Simmons98,Yoon99,Papadakis99}. These systems are also known to exhibit an in-plane transport anisotropy\cite{Adavies91,Heremans92,Heremans94b}; the mobility along the [$\bar{2}$33] direction is typically larger than along [01$\bar{1}$] by a factor of 2 to 5. Here we report the observation of commensurability oscillations (COs) revealing a surprisingly large anisotropy in the amplitude of the COs measured along these two directions in this system. We recall that COs refer to the oscillations in the magnetoresistance of a 2D carrier system whose density has a lateral periodic modulation\cite{Weiss89,Winkler89}. In a simple, semiclassical model\cite{Beenakker89}, the oscillations can be related to the classical cyclotron orbit diameter of the carriers becoming commensurate with a multiple integer of the period a of the potential. The anisotropy of the COs we observe in the 2DHSs is too large to be explained by a simple argument based on mobility anisotropy.

Our samples were grown by molecular-beam epitaxy (MBE) on GaAs (311)A substrates, and have low-temperature mobility in the range of 1.7$\times$10$^{5}$ cm$^{2}$/V s to 1.6$\times$10$^{6}$ cm$^{2}$/V s. The sample structure is shown schematically in Fig. 1. Arrays of 1000{\AA}-thick PMMA strips were defined perpendicular to the two arms of an L-shaped Hall bar pattern which was made by electron-beam lithography and 5000{\AA}-deep wet etching. The two arms of the Hall bar were aligned along the [$\bar{2}$33] and [01$\bar{1}$] directions. The periods of PMMA strips range from 2000{\AA} to 3000{\AA} and their widths equal half of their periods. A front gate was made by evaporating 50{\AA} of Ti and 2000{\AA} of Au. We measured the longitudinal magnetoresistance as a function of perpendicular magnetic field $B$, at T $\approx$ 0.5K using a standard low-frequency lock-in technique.

\begin{figure}[b]
\centerline{
\epsfxsize=3.0in
\epsfbox{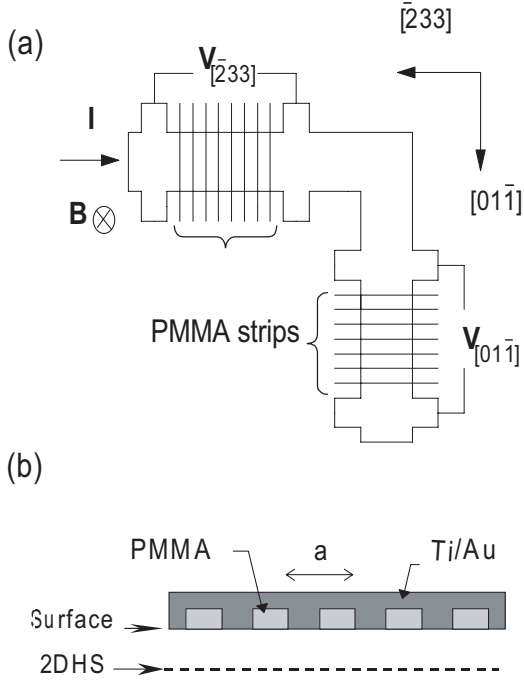}
}
\vspace{0.5cm}
\caption{(a) L-shaped Hall bar pattern for studying COs in a 2DHS on the GaAs (311)A surface. Periodic potentials along the  [$\bar{2}$33] and [01$\bar{1}$] directions are realized by e-beam defined PMMA strips and evaporated gates. (b) Schematic side view of the sample structure. Here {\it a\/} is the period of PMMA strips.}
\label{Fig1}
\end{figure}

\begin{figure}[t]
\centerline{
\epsfxsize=3.0in
\epsfbox{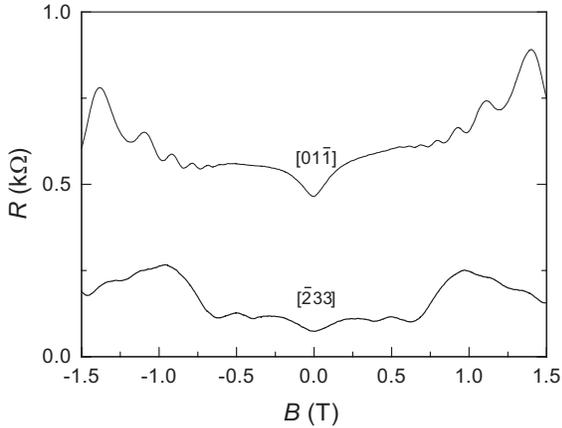}
}
\vspace{0.5cm}
\caption{Magnetoresistance measured in a triangular well 2D hole sample. Strong COs are observed in the [$\bar{2}$33] direction (thick trace), but no COs are observed in the [01$\bar{1}$] direction (thin trace). The high-frequency oscillations observed above $B\sim$ 0.5T are the Shubnikov-de Haas oscillations.}
\label{Fig2}
\end{figure}

\begin{figure}
\centerline{
\epsfxsize=3.0in
\epsfbox{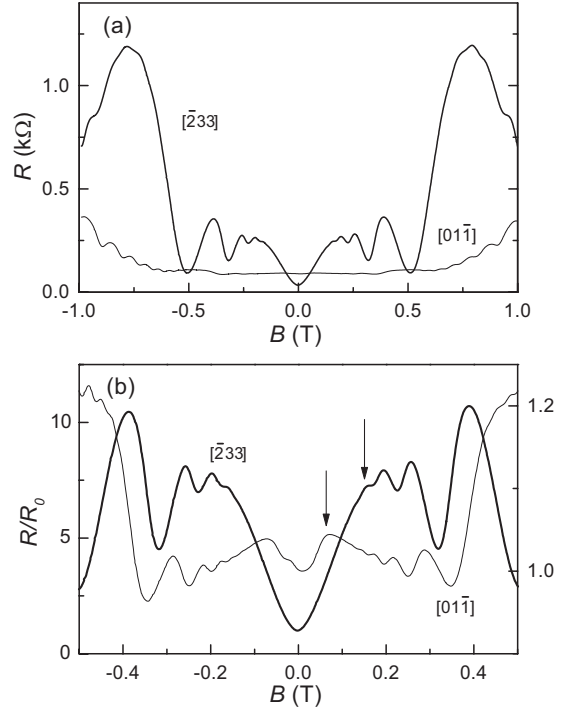}
}
\vspace{0.5cm}
\caption{(a)Magnetoresistance measured in a square well 2D hole sample. We observe strong COs in the [$\bar{2}$33] direction (thick trace), while the [01$\bar{1}$] direction (thin trace) exhibits very weak COs. (b)The COs along [$\bar{2}$33] and [01$\bar{1}$] are normalized by dividing by $R_{0}$, the resistance at $B=$0, to emphasize the enormous ($\sim$100 times) difference in their amplitudes. The vertical arrows mark the positions of the critical field below which a positive magnetoresistance is seen.}
\label{Fig3}
\end{figure}

We studied 2DHSs confined in both triangular and square quantum wells. The low-$B$ magnetoresistances measured along the [$\bar{2}$33] and [01$\bar{1}$] directions are represented by $R_{[\bar{2}33]}$ and $R_{[01\bar{1}]}$, respectively. Both $R_{[\bar{2}33]}$ and $R_{[01\bar{1}]}$ are measured with the current running perpendicular to the direction of the PMMA strips (see Fig. \ref{Fig1}). We find an enormous amplitude anisotropy in the COs measured along the [$\bar{2}$33] and [01$\bar{1}$] directions. In the triangular well sample, as shown in Fig. \ref{Fig2}, we observe strong COs along [$\bar{2}$33] while essentially no COs are visible along [01$\bar{1}$]. Significant anisotropy in COs is also observed in the square well sample. As shown in Fig. \ref{Fig3}(a), $R_{[\bar{2}33]}$ exhibits strong COs for 0.1 $<|B|<$ 1T while $R_{[01\bar{1}]}$ shows much weaker oscillations. In Fig. \ref{Fig3}(b), where $R_{[\bar{2}33]}$ and $R_{[01\bar{1}]}$ are normalized by $R_{[\bar{2}33]}$ ($B=$0) and $R_{[01\bar{1}]}$ ($B=$0), respectively, shows that $R_{[01\bar{1}]}$ does have some oscillating features but their amplitude is about 100 times smaller than along [$\bar{2}$33].

Surface corrugations, consisting of alternating GaAs and AlGaAs channels along [$\bar{2}$33], are present in MBE-grown GaAs (311)A surfaces\cite{Notzel93,Wassermeier95}. Formation of these channels has previously been reported\cite{Adavies91,Heremans92,Heremans94b} to lead to enhanced interface roughness scattering along [01$\bar{1}$], causing the mobility along [01$\bar{1}$] to be smaller than along [$\bar{2}$33]. Previous studies also indicate\cite{Adavies91,Heremans92,Heremans94b} that the mobilities along [$\bar{2}$33] and [01$\bar{1}$] differ only by a factor of 2 to 5. This difference is obviously far less than the nearly 100 times difference observed in the amplitudes of COs along [$\bar{2}$33] and [01$\bar{1}$] in our square well sample, and cannot explain the dramatic anisotropy observed in COs.

\begin{figure}[t]
\centerline{
\epsfxsize=3.0in
\epsfbox{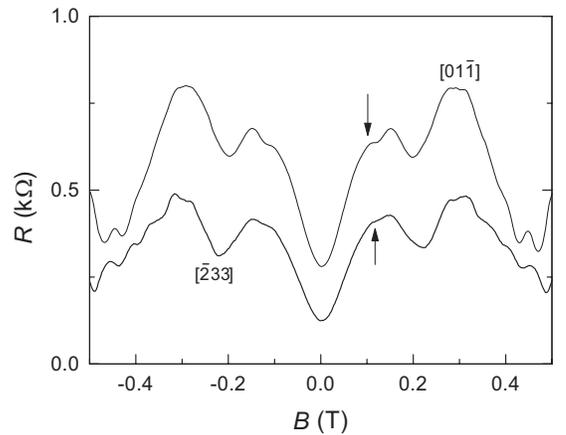}
}
\vspace{0.5cm}
\caption{Magnetoresistance measured in a 2D electron sample. The data along both [$\bar{2}$33] and [01$\bar{1}$] directions show COs with similar amplitudes. The vertical arrows also mark the positions of the critical field.}
\label{Fig4}
\end{figure}

We also studied the COs in a two-dimensional {\it electron\/} system (2DES) at the GaAs/AlGaAs (311)A interface. The 2DES was grown under similar MBE growth conditions: The substrate temperature during the growth of the GaAs/AlGaAs interface and the AlGaAs spacer was kept the same as in the 2DHS case and was reduced only when the dopants were introduced \cite{Agawa94}. The interface morphology of the 2DES sample should therefore closely resemble that of the 2DHS samples. The mobility measured in this sample along [$\bar{2}$33] is about 2 times greater than along [01$\bar{1}$]. The COs measured along [$\bar{2}$33] and [01$\bar{1}$] in the 2DES are shown in Fig. \ref{Fig4}. It is clear that the amplitude of the COs in the 2DES case differ only by a factor of about 2, much less than the nearly 100 times difference observed in the 2DHSs. From the positions of the critical field (marked by the vertical arrows in Fig. \ref{Fig4}) for the positive magnetoresistances measured along [$\bar{2}$33] and [01$\bar{1}$], we estimate that the magnitude of the potential modulation is about the same along these two directions. These values are reasonably consistent with the amplitude difference of COs observed along [$\bar{2}$33] and [01$\bar{1}$] in Fig. \ref{Fig4}.

To understand the data more quantitatively, it is useful to examine the strength of the potential modulation along [$\bar{2}$33] and [01$\bar{1}$]. Beton {\it et al\/}\cite{Beton91} show that 2D carriers subjected to a one-dimensional periodic potential exhibit a low-field positive magnetoresistance. This is because the carriers, affected by the electrical force generated by the periodic potential, stream along the equipotentials (perpendicular to the current flow), and lead to an increase in the resistance. This positive magnetoresistance extends up to a critical $B$ field (marked by the vertical arrows in Fig. \ref{Fig3}(b)) above which the magnetic force is greater than the electrical force and the streaming orbits are suppressed. Therefore, the critical $B$ field can be a measure of the strength of the potential modulation. From this model, we estimate that the strength of the potential modulation along [$\bar{2}$33] is about twice larger than along [01$\bar{1}$] for the 2DHS sample. This is not sufficient to explain the remarkable COs amplitude anisotropy found in the 2DHS. The modulation strength for the 2DES sample is about the same for the two directions. 

In summary, we investigated COs in 2DHSs and a 2DES grown on GaAs (311)A substrates under similar MBE growth conditions and found an anomalous anisotropy in the COs' amplitudes measured along [$\bar{2}$33] and [01$\bar{1}$] in the 2DHSs, but not in the 2DES. Our study suggests that the anomalous anisotropy is intrinsic to the 2DHSs grown on GaAs (311)A substrates.

This work was supported by the NSF and the ARO.


\vspace{0.7cm}

\end{document}